\documentclass[fleqn,10pt]{wlscirep}
\usepackage[utf8]{inputenc}
\usepackage[T1]{fontenc}
\usepackage{amsmath}
\usepackage{tabularx}
\usepackage{url}
\title{Flickering candle flames \\ and their collective behavior}

\author[1]{Attila Gergely}
\author[1]{Bulcs\'u S\'andor}
\author[2]{Csaba Paizs}
\author[2]{Robert T\"ot\"os}
\author[1,*]{Zolt\'an N\'eda}
\affil[1]{Department of Physics, Babe\c{s}-Bolyai University, Kog\u{a}lniceanu street nr. 1, RO-400084 Cluj-Napoca, Romania}
\affil[2]{Biocatalysis and Biotransformation Research Centre, Faculty of Chemistry and Chemical Engineering, Babe\c{s}-Bolyai University,
Arany J\'anos street nr. 11, RO-400029, Cluj-Napoca, Romania}

\affil[*]{zneda@phys.ubbcluj.ro}

\newpage

\begin{abstract}
Oscillation and collective behavior of diffusion flames is a fascinating phenomena. Considering candle bundles with different sizes in variable oxygen concentration, 
the flickering dynamics of the flames are experimentally and theoretically investigated. Trends for the flickering frequency as a function of the candle number in the bundle
and oxygen concentration is revealed for various topologies of the candles packing. The collective behavior of the flames as a function of their separation distance  is 
studied by measuring an appropriate synchronization order parameter and trough the common oscillation frequency. In agreement with previous results we find a discontinuous phase transition between an in-phase synchronized state at small separation distance and a counter-phase synchronized state at larger separation distances. 
A previously used dynamical model is modified in order to accommodate our experimental findings.
\end{abstract}

\begin{document}

\flushbottom
\maketitle
%
%
\thispagestyle{empty}

\section*{Introduction}
Oscillation of diffusion flames is an intriguing phenomena, known for a long time \cite{Chamberlin1948}. Controlling these instabilities are important in applications where one needs to stabilize the flame. Many studies have been conducted to investigate the effect of various parameters on the oscillation frequency and to explain the 
cause of the oscillation. It has been shown that the oscillation frequency scales as a power law with the nozzle diameter \cite{Huang1999}, external pressure \cite{DUROX1997}, and gravitational acceleration \cite{Durox1995} (centrifuge experiments). Early studies suggested that the oscillation frequency does not depend on the fuel outflow velocity, however, measurements for a wider ranges observed a power law dependence for this as well \cite{Zeng2013}.  In order to explain the observed oscillations, either an approach based on hydrodynamic instabilities \cite{Yuan1994} or a dynamical system approach based on the mechanism of the chemical reaction on the surface of the flame \cite{Kitahata2009} was considered.

Recently this problem returned in the focus of researchers and a number of studies \cite{Kitahata2009,Ghosh2010,Forrester2015,Okamoto2016,Manoj2018,Chen2019,Yang2019} have been published on the oscillation of candle flames by considering different setups. Beside the oscillation of the candle flames, the collective behavior of an ensembles of oscillating flames close enough to each other was also investigated \cite{Kitahata2009,Okamoto2016}. For short distances between the oscillating flames in-phase synchronization and for larger distances anti-phase synchronization was observed. By further increasing the distance between the 
oscillating flames their flickering becomes uncorrelated.  Synchronization was observed recently also for methane diffusion flames \cite{Fujisawa2020}. 

In the present work we reconsider both experimentally and theoretically the oscillation of the candle flames and their collective behavior.  We report on experimental results that lead to modifications in the original model proposed by Kitahata et. al \cite{Kitahata2009}. More precisely, in the case of bundles with different topologies we examine the effect of the bundle size on the oscillation frequency in a much wider range than in the previous studies \cite{Chen2019}, and we examine the effect of the oxygen concentration on the oscillation. We reproduce the results known so far for the synchronization of two bundles \cite{Kitahata2009,Chen2019} and report inconsistencies with  the coupling mechanism proposed earlier \cite{Kitahata2009}.  The dynamical system approach presented in the work of Kitahata et. al \cite{Kitahata2009} is revised by changing the equations so that they lead to realistic trends as a function of the system parameters both for the oscillation of the flame in one candle bundle and for their collective behaviour as a function of the separation distance. 

The rest of the paper is organized in the following manner. First, we present the dynamical model used previously to understand the oscillation of candle flames and their collective behavior. We describe our experimental methods and results arguing for the need of revised dynamical equations.  A modified dynamical model is then offered that allows a good understanding of the observed oscillation and synchronization phenomena in agreement with the experimental results. Finally, we summarize our findings and discuss on the applicability of the results.    


\section*{\label{sec:model} Modeling framework for candle flame oscillation and their collective behavior}

We start from the classical dynamical system approach proposed by the Japanese group \cite{Kitahata2009}.  It consists of two coupled first-order differential equations for the main quantities that are considered to be 
decisive in understanding the flame's dynamics: temperature, $T$, and oxygen concentration, $n$, in the flame:  
\begin{equation}
\begin{aligned}
C \frac{dT}{dt} &= \omega_1 \left[-h\cdot (T-T_0)+\beta \cdot n\cdot a \cdot e^{-\frac{E}{RT}}\right]-\sigma \cdot T^4\\
\frac{dn}{dt} &= \omega_2\left[k\cdot (n_0-n)-a\cdot  n\cdot e^{-\frac{E}{RT}}\right]
\end{aligned}
\label{eq:japan_eredeti}
\end{equation}

The first equation  describes the energy conservation: thermal energy of the system is diminished by heat loss through
convection (first term on the right side) and  radiation (third term), and it is increased due to heat production by
burning. The heat production rate depends on a Boltzmann factor governed by a chemical activation barrier $E$,  it is linearly proportional with the oxygen concentration and the fuel (paraffin) supply rate. We denoted by $C$ the heat capacity of the system, $\omega_1$ is a characteristic time-scale (frequency for the
process), $h$ characterizes the heat conduction, $T_0$ is the external temperature of the environment, $\beta$ is a proportionality factor, $a$ is the fuel consumption rate,  $\sigma$ is the heat radiation coefficient and $R$ is the ideal gas constant. 

The second equation describes the oxygen balance according to which the amount of oxygen in the flame increases due to a flux from the exterior governed by a classical transport phenomena with the proportionality constant $k$ and it is decreased due to oxygen consumption by burning.  We denoted by $\omega_2$ a characteristic frequency (time-scale) and by $n_0$ the oxygen concentration in the external atmosphere.  

\begin{figure}
\includegraphics[width=0.45\linewidth]{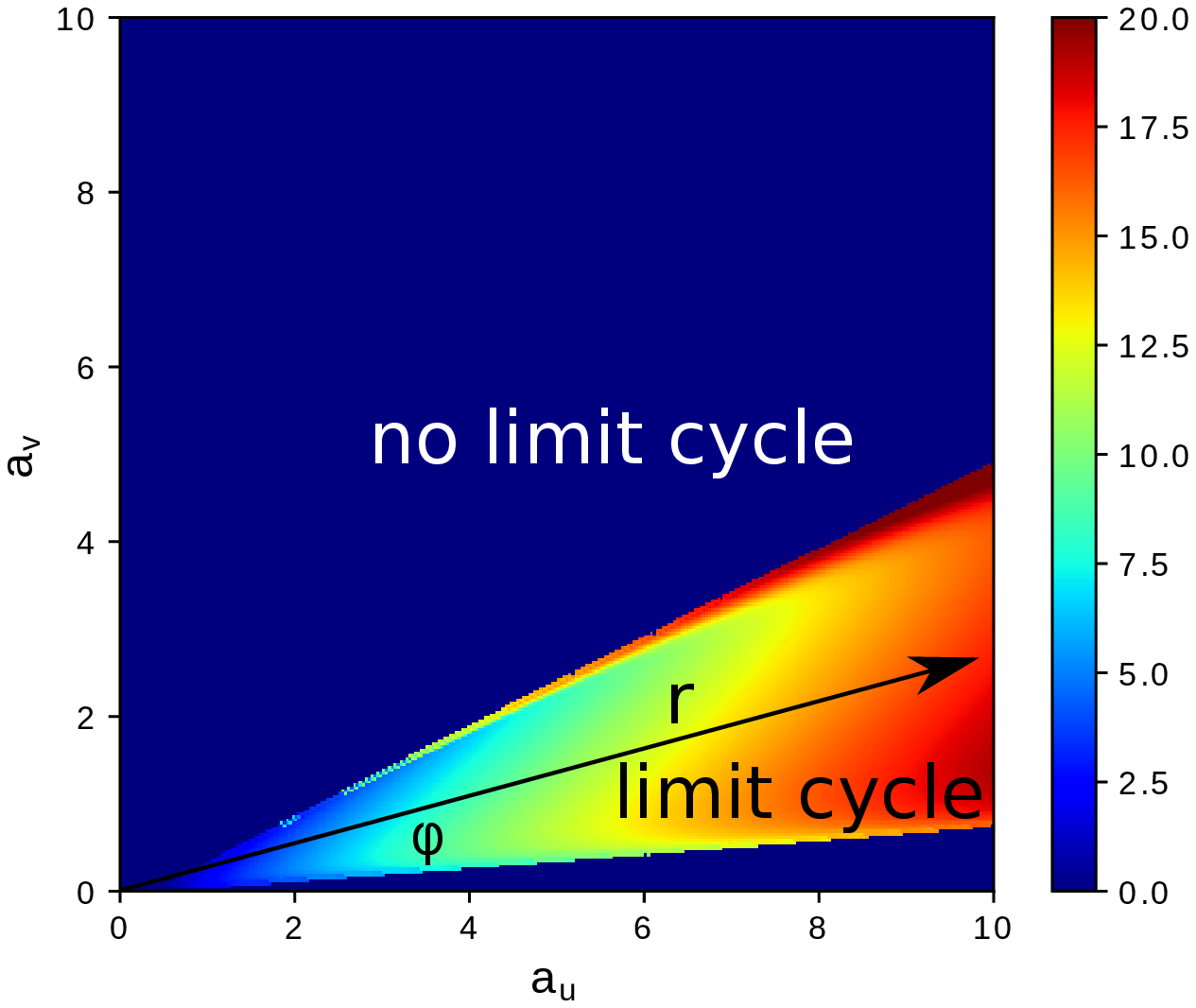}
\includegraphics[width=0.55\linewidth]{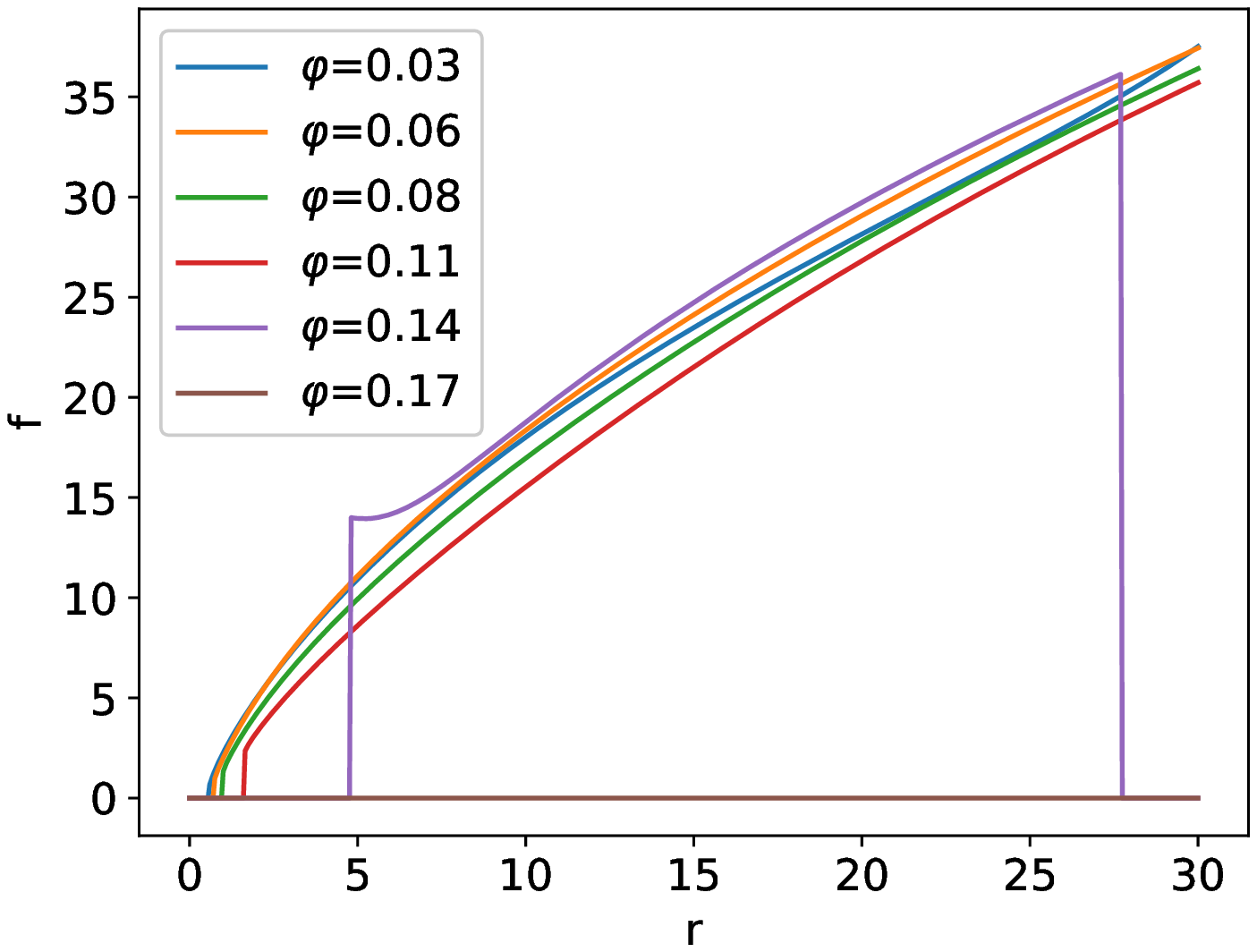}
\caption{The figure from the left shows with a color-code the 
frequency of the limit-cycle oscillations in the $a_u$, $a_v$ parameters space, obtained from equations (\ref{eq:sadim1}) using the parameters  $\varepsilon = {10}^{-3}$, $c = 5.1$, $\sigma_0=1$. In the dark-blue region there is no limit-cycle. For increasing the number of candles in the bundle (increasing the value of $a$) one moves in the direction of a line as indicated in the figure by the black arrow. The figure on the right side illustrates the variation of the frequency for the limit-cycle 
oscillations when the slope $\varphi$ is constant and $r$ is increasing. Here $r=\sqrt{a_u^2+a_v^2}$ and 
$\varphi=\arctan{(a_v/a_u)}$. Depending on the value of $\varphi$ one might not get any oscillations ($\varphi>0.17$rad for the indicated directions).} 
\label{fig:hat_c}
\end{figure}

By introducing the following non-dimensional parameters,

\begin{equation}
\begin{split}
\tau=t\,\omega_2\, k\,, \qquad c=\frac{E}{RT_0}\, \qquad \\
u=c\,\frac{T-T_0}{T_0}\,, \qquad
v=\frac{n}{n_0}\, \qquad \\
a_u=\frac{\beta \,c\, n_0 \,a }{T_0 \,h}e^{-c}\,, \qquad
a_v=\frac{a}{k}e^{-c}\qquad \\
\varepsilon^{-1}=\frac{h\omega_1}{C\,k\,\omega_2}\,, \qquad \sigma_0=\frac{\sigma \, c\,T_0^3}{C\, \omega_2 \,k} \qquad
\end{split}
\label{eq:padim}
\end{equation}

the dynamical equations can be written also in a non-dimensional form:

\begin{equation}
\begin{aligned}
\frac{du}{d\tau} &= \frac{1}{\varepsilon}\left[-u+a_uve^{\left(\frac{uc}{c+u}\right)}\right]-\sigma_0{\left(1+\frac{u}{c}\right)}^4\\
\frac{dv}{d\tau} &= 1-v-a_vve^{\left(\frac{uc}{u+c}\right)}
\end{aligned}
\label{eq:sadim1}
\end{equation}
This coupled first-order differential equation system can be numerically integrated, and it's behavior as
a function of the $a_u$, $a_v$, $\epsilon$, $c$ and $\sigma_0$ parameters can be studied. The flame diameter or
the number of candles, $N$, in the bundle governs the value of the $a$ parameter (and therefore the values of $a_u$ and $a_v$), which should increase with the size of the flame. By fixing the $\epsilon$, $c$ and $\sigma_0$ parameters one can study the behavior of the system in the $a_u-a_v$ parameter plane and identify the region where there is limit cycle, i.e. the flame oscillates. For realistic parameters ($\epsilon=0.001$, $c=5.1$ and $\sigma_0=1$) the systems behavior is illustrated in Figure \ref{fig:hat_c}.  The color code in the figure illustrates the oscillation frequency of the flame. In the violet 
region the system has a fix point and no limit-cycle, therefore the flame is stable. Depending thus on the 
$a_u$ and $a_v$ parameters one can observe either a stable flame or an oscillation. Increasing the flame diameter 
(or considering candle bundles with more and more candles) one increases the $a$ parameter and proportionally the
values of $a_u$ and $a_v$, going away from the origin on a straight line trajectory in the $a_u-a_v$ parameter space (see 
Figure \ref{fig:hat_c}). This behavior suggests that flickering (oscillation) starts only after $a$ exceeds a given value, and the oscillation frequency should increase with increasing the value of $a$, hence  with the number of candles in the 
bundle. The behavior is similar also for other $\epsilon$, $c$ and $\sigma_0$ values. 

Kitahata et. al suggested that the in-phase and anti-phase synchronization of two candle bundles can be understood by coupling the dynamical equations (\ref{eq:japan_eredeti}) for two systems through the thermal radiation term. For two
identical bundles $i\ne j \in\{1,2\}$ separated at distance $x$ the coupling writes as:

\begin{equation}
\begin{aligned}
 \frac{dT_i}{dt} &= \frac{\omega_1}{C} \left[h\cdot (T_0-T_i)+\beta \cdot n_i\cdot a \cdot e^{-\frac{E}{RT_i}}\right]-\frac{\sigma}{C} \cdot T_i^4+\frac{\sigma}{C} \frac{\mu}{x^2}T_j^4\\
\frac{dn_i}{dt} &= \omega_2\left[k\cdot (n_0-n_i)-a\cdot  n_i\cdot e^{-\frac{E}{RT_i}}\right]
\end{aligned}
\label{eq:japan_coupled}
\end{equation}

The authors argue \cite{Kitahata2009} that the above coupled system is able to reproduce for increasing 
separation distance $x$, the in-phase and anti-phase synchrony as well. Rewriting equations (\ref{eq:japan_coupled}) again in the non-dimensional form and using the same parameters as in (\ref{eq:padim}) , the authors state that for the parameters 
$\epsilon=0.001$, $a_u=a_v=3.7$, $c=5$ and $\sigma_0=1$ for $\mu/x^2=0.5$ one gets in-phase synchronization of the flames, while for $\mu/x^2=0.01$ anti-phase synchronization is achieved. Unfortunately one cannot reproduce the 
results mentioned above for the parameters given in the study \cite{Kitahata2009}. However, for $a_v=1$ and keeping all the other parameters as mentioned above, one can reproduce the desired collective behavior although the obtained oscillation frequency will be very different from the one suggested by Kitahata et. al. It should be also kept in mind that this model for the fixed parameter sets will lead to anti-phase sync only for very restricted initial conditions. With other words the basin of attraction for obtaining the anti-phase synchronization is relatively small. 

Although the above presented dynamical model is a simple one and captures the main characteristics of the 
investigated phenomena, it proves to be in disagreement with our experimental results presented in the followings. 

\section*{\label{Kiserleti_r}Experimental part}
\subsection*{\label{berendezesek_m} Equipments and methods}

\begin{figure}
 \centering
    \includegraphics[width=0.45\textwidth]{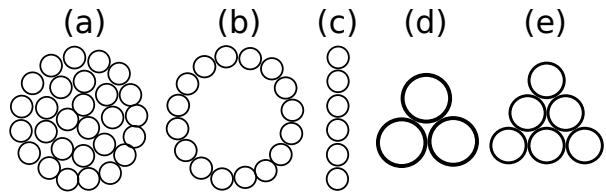}
    \caption{Arrangements of the candle bundles used in the experiments. Arrangement (a) is referred to as a compact arrangement, (b) as a hollow arrangement, (c) is a linear arrangement and (d) and (e) as a triangular arrangement. The topologies from (a) - (d) were used in studying the oscillation frequencies of a single bundle, while the topologies (d) and (e) were considered for investigating the collective behavior of the flame from two bundles.}
 \label{jukas}
\end{figure}

Since the flame of a single candle usually does not exhibit oscillations, candle bundles with various sizes and in different topologies were built. We used 8mm diameter and 7cm long candles arranged in the topologies depicted in  Figure \ref{jukas}. The experiments were performed at room temperature under normal atmosphere and in different oxygen+nitrogen mixtures as well. For the compact and hollow arrangements, (a) and (b) respectively,  the effect of the bundle size on the oscillation frequency was examined for a wide range of the candles number. The effect of the oxygen concentration was studied qualitatively for a single candle and  in more detail for the triangular arrangement with 3 candles (Figure \ref{jukas}d). The collective behavior of two bundles was studied as a function of their distance using the triangular arrangements (Figure \ref{jukas}d and \ref{jukas}e) for the bundles. 

The oscillations of the candle flames and the emerging collective behavior was studied experimentally both with a TroubleShooter high-speed camera and with an Atmega 16 chip-based photoresistor device designed by us. 

\begin{figure}
 \centering
    \includegraphics[width=0.55\textwidth]{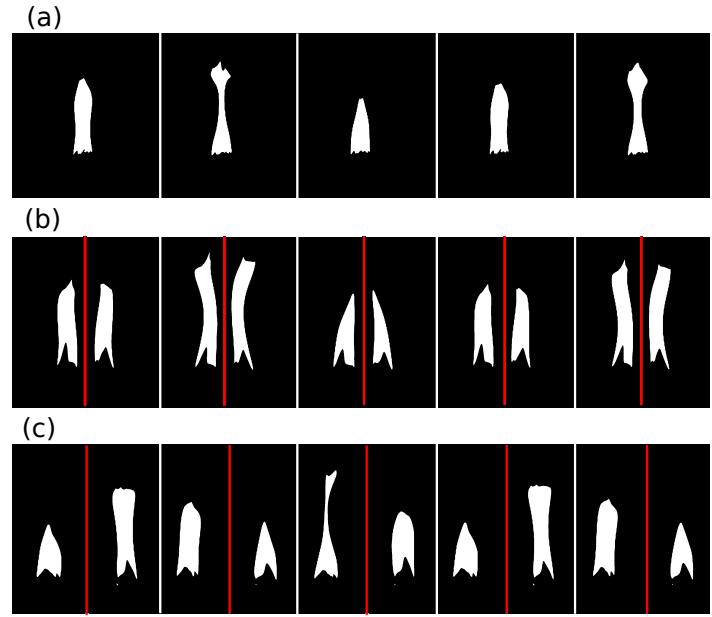}
    \caption{Recorded frames with the TroubleShooter high-speed camera and processed with the Otsu method. Figure (a) presents frames for the oscillation of one bundle. Figure (b) are  frames for the in-phase synchronized dynamics of two bundles, and Figure (c) presents frames for the anti-phase synchronized dynamics of two bundles.  }
 \label{frames}
\end{figure}

The measurement with the high-speed camera was performed as follows: from 2000 up to 8000 frames were recorded with 250fps or 1000fps of the flickering (as an example see Figure \ref{frames}), and then the pixels belonging to the flame on each frame were identified using the Otsu method \cite{Otsu:2020}. For a single bundle the time series were constructed for the number of pixels identified in the flame, giving a gauge for the flames intensity. For the collective behavior of two bundles the identified pixels for the flames were first separated in two parts, belonging to each flame in part, as it is depicted by the red line in the figure.

The Atmega 16 chip is connected to a one-step voltage divider to convert the brightness into a voltage value. The voltage value is measured by the analog input of the microchip, and the measured result is transmitted to the computer via asynchronous communication. The sampling frequency is 2kHz.

\begin{figure}
    \centering
    \includegraphics[width=0.7\textwidth]{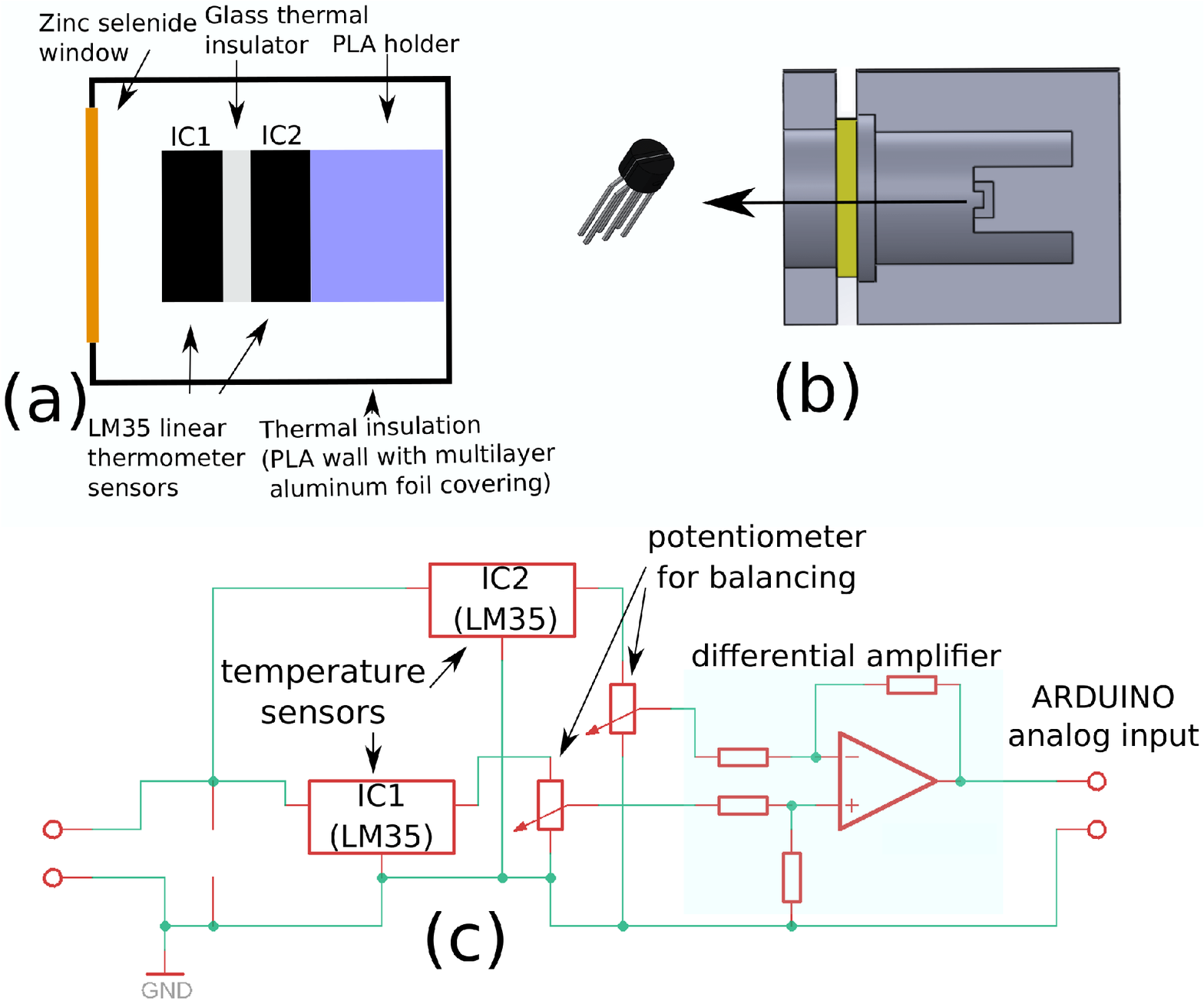}
     \includegraphics[width=0.6\textwidth]{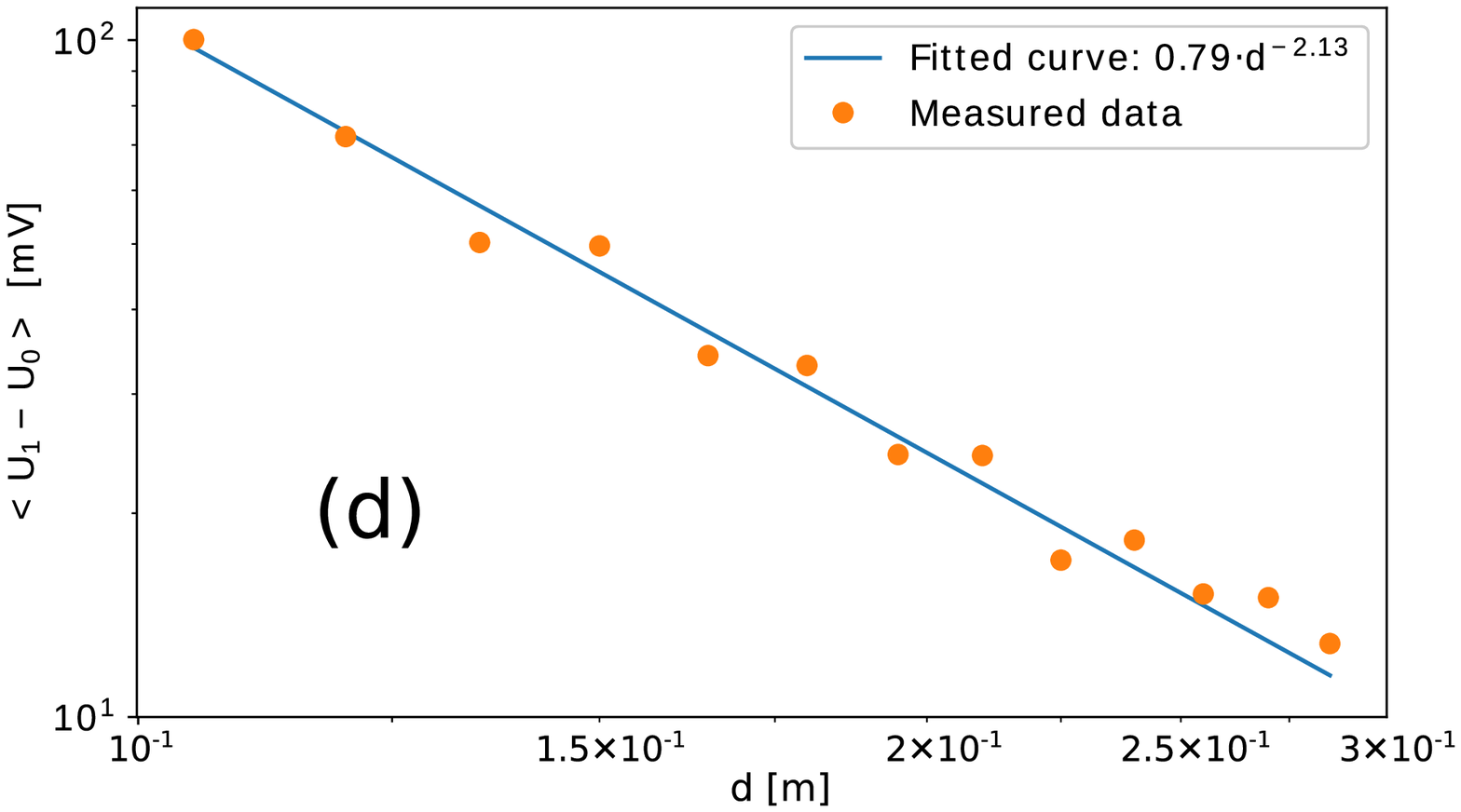}
 \caption{Subfigure (a) shows the scheme of thermal radiation measuring equipment. Thermal radiation passes through a zinc selenide window to the IC1 sensor where it is absorbed. The absorbed heat passes through the glass plate and the sensor IC2 into the housing of the device, it can be shown that if the incoming thermal radiant energy is constant, the temperature difference between the sensors is proportional to the incoming thermal power. Subfigure (b) shows a section and a side view of a 3D model of the device. Subfigure (c) shows the circuit diagram of the device. The measurement consists of two steps: we measure the voltage $U_0$ given by the device when the input is covered and also when the thermal radiation reaches the sensor $U_1$ the voltage difference $U_1-U_0$ is the measurement result (proportional to the incoming thermal power). Subfigure (d) is the calibration curve for the thermal radiation measuring device. The figure illustrates the signal given by the measuring device as a function of the distance measured from a 50W halogen bulb. From the linear fit, we learn that the resulting voltage difference is proportional to the inverse of the square of the distance. Please note the double-logarithmic scale.  }
    \label{fig:sugarzasmero}
\end{figure}

For measuring the thermal radiation of the flames we constructed an original device. Our sensor is schematically presented in Figure  \ref{fig:sugarzasmero} (a) and (b). The skeleton of the device was 3D printed. The thermal radiation enters in the detector through a ZnSe window to allow a wide transmission band even in the deep infrared range. Two sensitive temperature sensors (IC1 and IC2), divided by an isolator glass plate, are placed one after the other as it is shown in the figure. The incoming radiation produces a temperature gradient in the glass plate, which equilibrates in time for a constant room temperature. This equilibrium value is proportional with the incoming radiation flux and it was measured with a sensitive setup. The circuit diagram of the measuring device is shown on Figure \ref{fig:sugarzasmero}(c).  It is designed for amplifying the small differences measured by IC1 and IC2 and also to compensate their manufacturing differences. The device was calibrated in such manner that for zero thermal radiation it gives no voltage. By using the thermal radiation of a halogen bulb placed at different distances, $d$, from the detector we proved that the device gives a linear response of the detected voltage as a function of the intensity of the thermal radiation.   In  Figure \ref{fig:sugarzasmero}d we illustrate on a log-log scale the voltage given by the sensor as a function of the distance from the radiation source, indicating a power-law dependence with exponent roughly $2$.  Taking into account that the intensity of the radiation decreases as a function of the distance as $1/d^2$, we can conclude that our sensor is linear in a good approximation.
 
In order to investigate the role of thermal radiation in the formation of the collective behavior of the candle bundles flame
we used also a pulsating radiation source with the same radiation properties as the candle flame. This was provided by a  halogen bulb of 50W which had the same  spectra as the one of the candles flame. More details about this experiment is given in the Supplementary Material, section A. 

\begin{figure}
    \centering
    \includegraphics[width=0.55\textwidth]{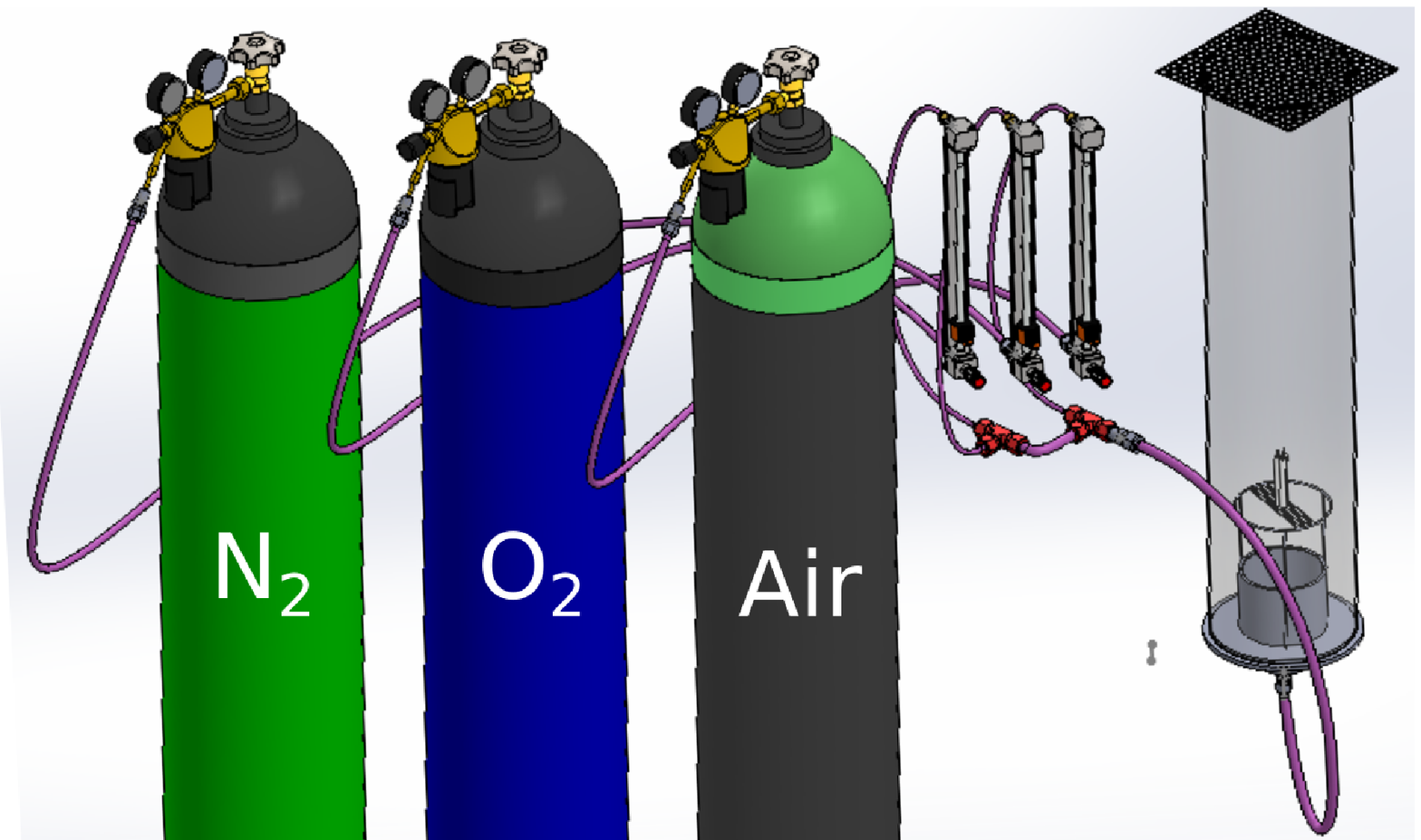}
    \caption{Experimental setup for investigating the influence of the oxygen concentration. As "air" we used synthetic air with the normal mixture of Oxygen (21$\%$) and Nitrogen (79$\%$).}
    \label{fig:oxigen}
\end{figure}

In the dynamical model proposed in the previous section, one can observe that the oxygen concentration in the environment is also an important parameter in the non-dimensional $a_u$ value. For studying also the effect of the air composition (oxygen concentration) we have constructed yet another experimental device (Figure \ref{fig:oxigen}). As it is illustrated, we 
mixed a composition of nitrogen, air and oxygen and by using fine regulators we could provide for the candle a controlled oxygen concentration. The candle bundle was placed in a tube of 15cm diameter and the gas entered below through honeycomb mash to achieve a uniform turbulence-free flow. The top of the pipe was covered with a fine metal mesh, which allowed the combustion products to escape and prevented the perturbing effect of the outside airflow.

\subsection*{\label{kiserleti_eredmenyek} Experimental results on the oscillation frequency}

The experiments were performed at room temperature at normal atmospheric pressure varying also the oxygen concentration.
Each experiment was repeated for at least 5 times, allowing thus to construct error-bars. Movies recorded with the high-speed camera 
for the oscillation of a single bundle in normal atmosphere and in controlled oxygen concentration can be consulted following our YouTube playlists 
\cite{Attila1,Attila2} dedicated to these experiments.

\begin{figure}
    \centering
    \includegraphics[width=0.455\textwidth]{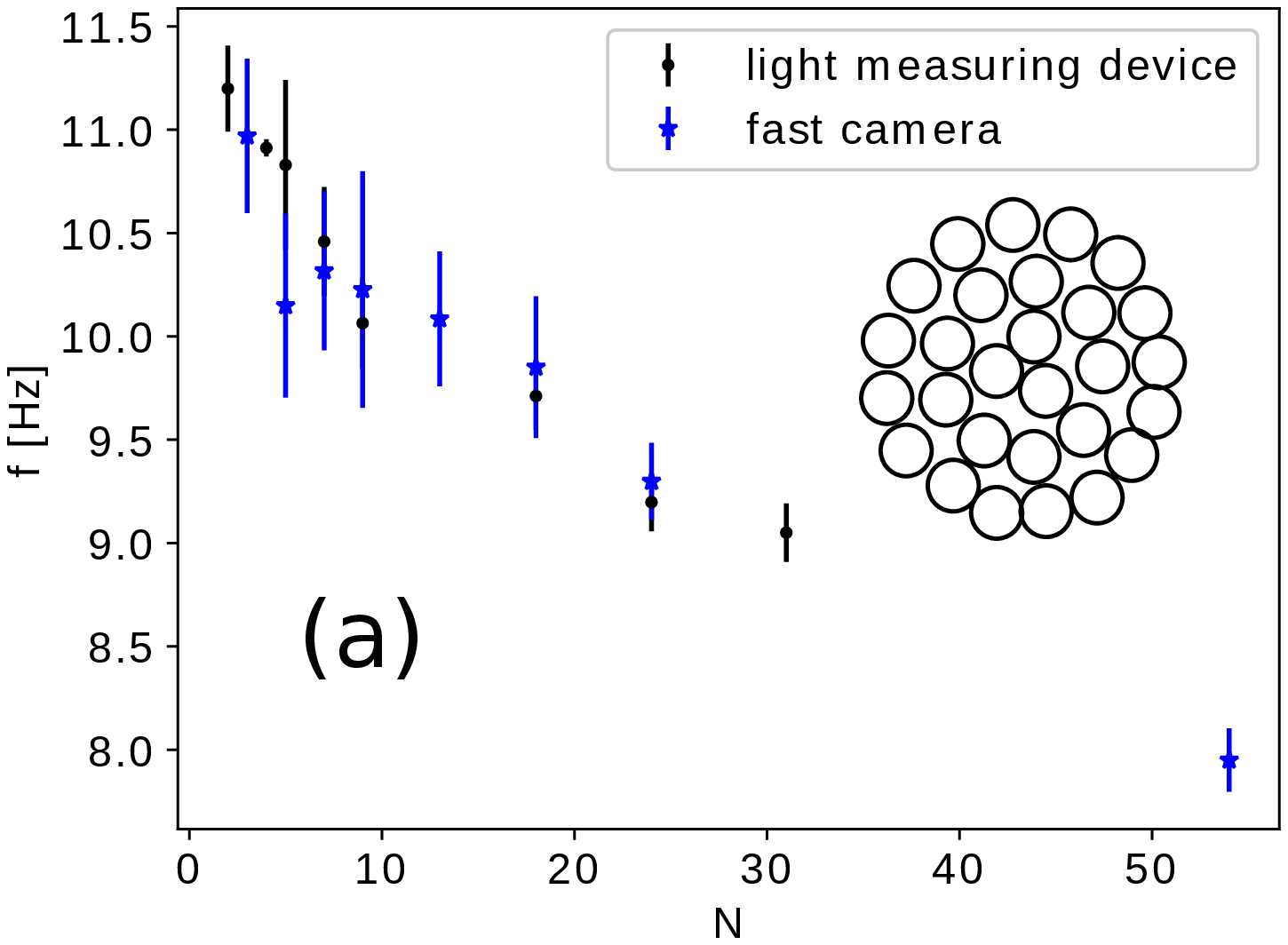}    
    \includegraphics[width=0.5\textwidth]{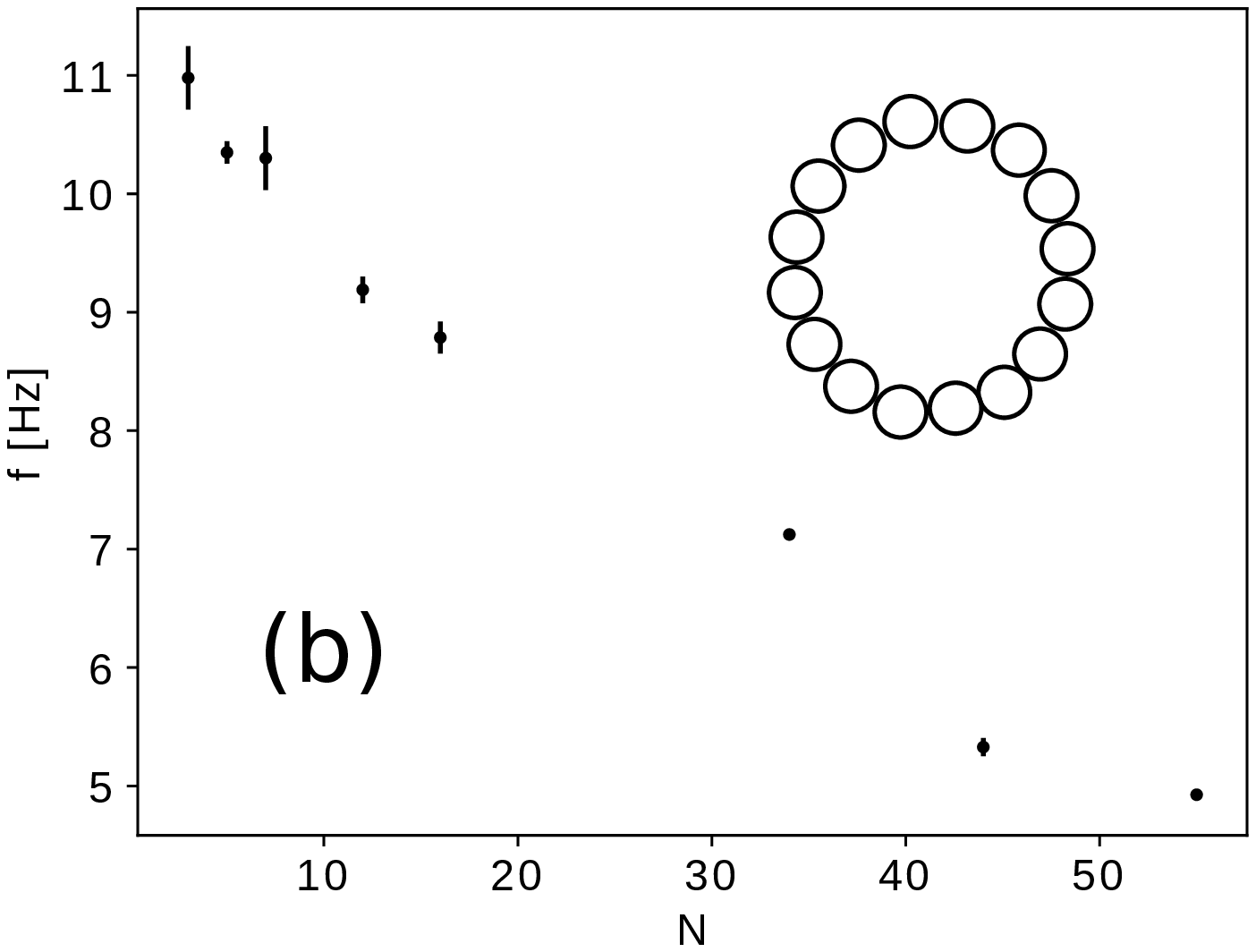}
    \includegraphics[width=0.5\textwidth]{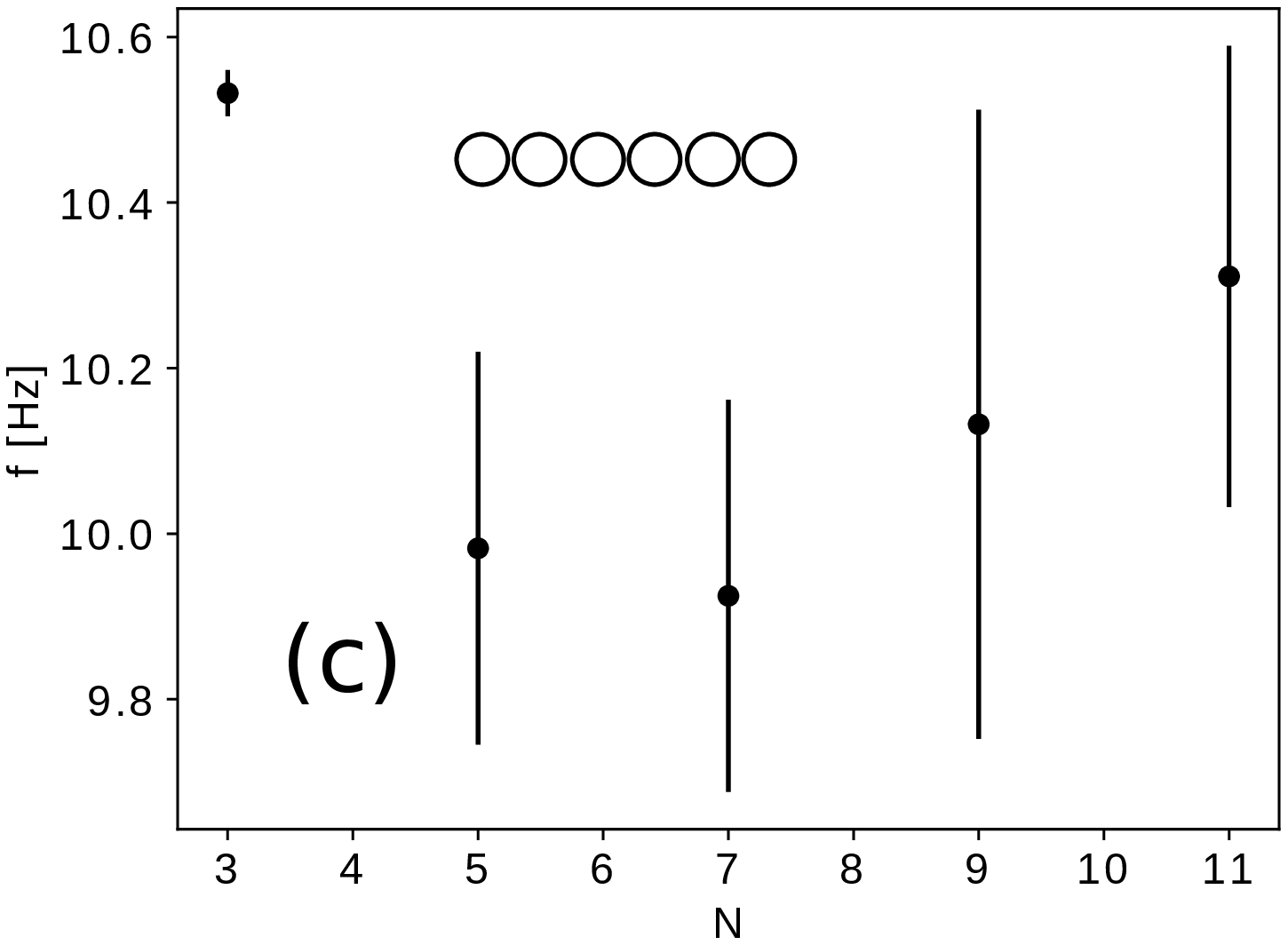}
    \caption{Frequency of flame flickering as a function of the number of candles, $N$, for various arrangements of the candles in the bundle. 
    In Figure (a), a compact arrangements are used for the candles, results are obtained both with the Atmega32 driven light measuring device and high-speed camera. 
    Figure (b) is for the hollow arrangement of the candles, measured with the light measuring device.  Figure (c) is for the
 linear arrangement of the candles, measured with the high-speed camera.}
    \label{fig:frequency}
\end{figure}

For the arrangements (a), (b) shown in Figure \ref{jukas}, the oscillation frequency, $f$, was examined for a wide size range of candle numbers, $N$, in the bundle. The measurement results are shown in Figures \ref{fig:frequency}a,b and c . For the compact and circular (hollow) topology a clear decreasing trend of the frequency as a function of the candle number is observed (Figure \ref{fig:frequency}a and Figure \ref{fig:frequency}b, respectively). For the linear arrangement no such monotonic trend is obtained (Figure \ref{fig:frequency}c). In this case for a higher number of candles ($N> 11$) the flame lost its compactness and therefore the oscillation frequency becomes ambiguous and we do not present results for such bundles \cite{Attila1}.   Our results outperforms the measurement from
\cite{Yang2019} by considering much larger bundles and various topologies. These results are in clear disagreement with 
the prediction of the dynamical model from (\ref{eq:japan_eredeti}), which suggests that the frequency should increase with the bundle size.

\begin{figure}[h]
    \centering
    \includegraphics[width=0.41\textwidth]{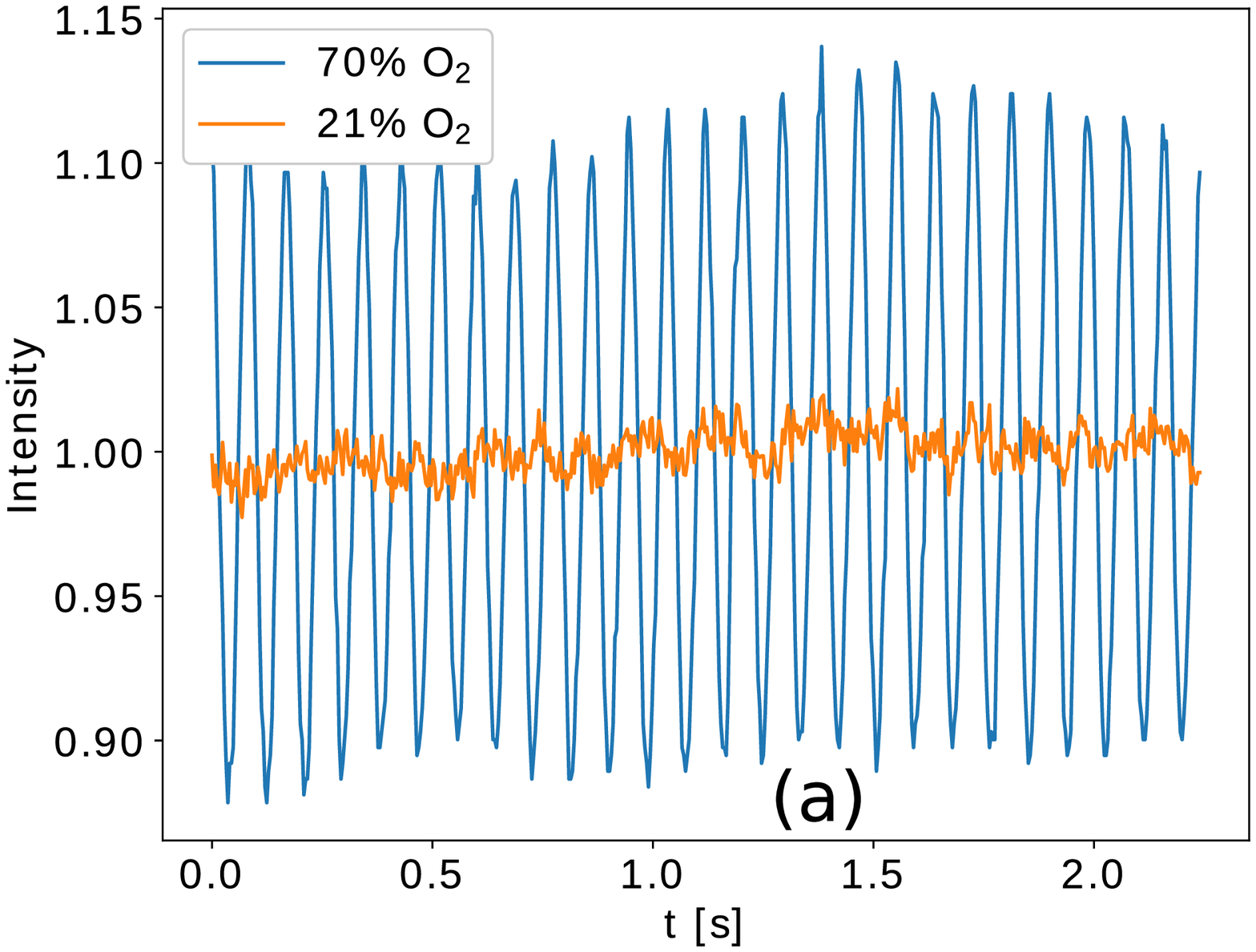}
    \includegraphics[width=0.55\textwidth]{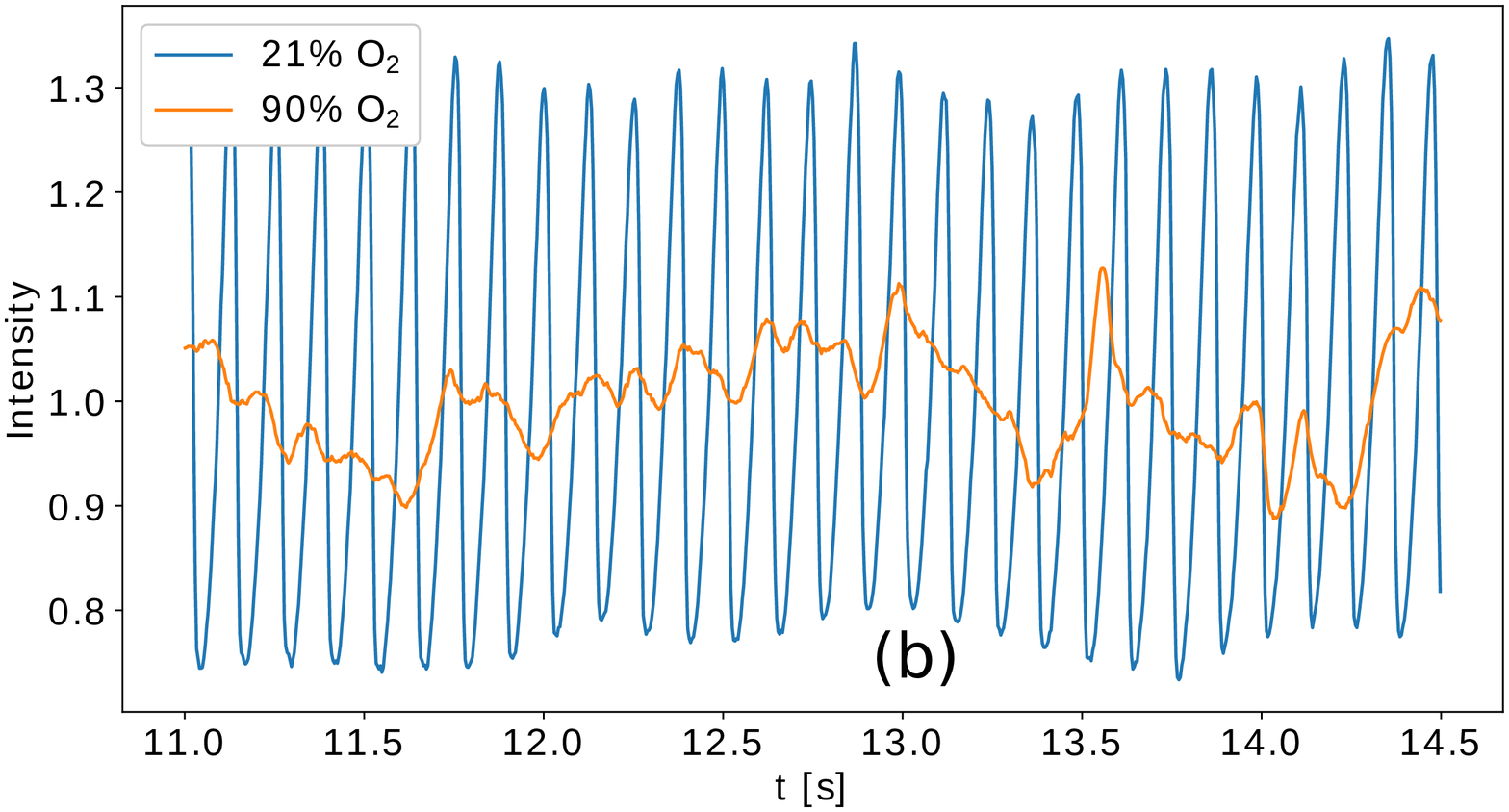}    
    \caption{Time series for the recorded flame intensity (values normalized to the mean) obtained with the high-speed camera and processed with the Otsu method. 
(a) A single candle flickering in an atmosphere containing 21\% and 70\% oxygen concentrations. (b) A bundle with 14 candles in a hollow arrangement at 21\% and 90\% oxygen concentrations. }
    \label{egyetlen_gy}
\end{figure}

The effect of the oxygen concentration, $n_0$, on the oscillation frequency was studied using the device described in the previous subsection. It was found that in contrast to what happens at normal oxygen concentration, the flame of a single candle also oscillates with a frequency of 11.54Hz when the oxygen concentration exceeds 70$\%$. A part of the time series obtained for intensity measurements recorded and processed for the 21\% and 70\% oxygen concentration is illustrated in Figure \ref{egyetlen_gy}a.  While the flame of a single candle at 21\% oxygen concentration shows no periodic oscillation, at 70\% oxygen concentration a clear periodicity is detected. 
The intensity was measured from the Otsu processed images.  First we determined the number of white pixels, $w_i$ in each frame $i$ and determined from these the average value $\langle w \rangle$. The intensity of the flame for each frame is a relative value, defined as the ratio $w_i/(\langle w \rangle)$. 

For the oscillation of a circular bundle containing 14 candles oscillating with a $f=8.05Hz$ frequency  in the presence of 21\% 
oxygen concentration we found that the oscillation disappears above 90$\%$ oxygen concentration. Figure \ref{egyetlen_gy}b shows two time series of the recorded intensity for the 14-element bundle at 21$\%$ and 90$\%$ oxygen concentrations, illustrating this phenomena. At 90$\%$ oxygen concentration one can observe a relatively 
weak fluctuation of the flame's relative intensity, a qualitatively different behaviour from the oscillations observed for 21$\%$ oxygen concentration. Unfortunately for other
bundle topology it was not possible to check this trend, since it would have required much larger number of candles in the bundle and/or higher oxygen concentration. Under such experimental conditions however the combustion would release an uncontrollably large amount of heat in the tube, becoming potentially dangerous and damaging the experimental setup. This indeed happened when we increased the oxygen concentration above 95\%. The only feasable alternative would have been to use the linear arrangements. In this case however the size of the tube would limit quickly the bundle size, and for small bundles again a dangerously high oxygen concentration would be necessary to achieve the same effect.

\begin{figure} [h]
    \centering
    \includegraphics[width=0.45\textwidth]{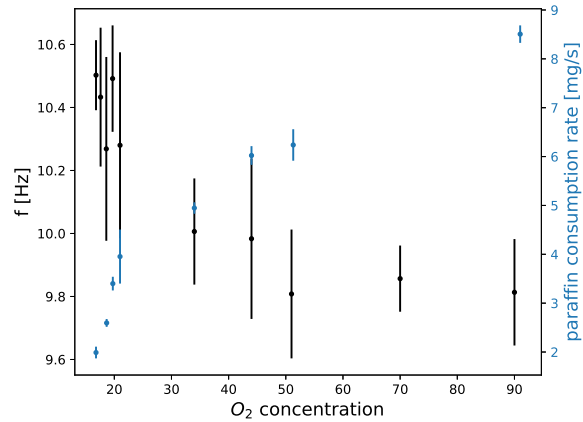}
    \caption{Oscillation frequency and the paraffin consumption rate as a function of oxygen concentration, $n_0$. Results for a triangular arrangement containing 3 candles. }
    \label{oxigen_konc}
\end{figure}

For different oxygen concentrations and using a triangular arrangement of 3 candles the oscillation frequency was 
studied with high-speed camera measurements.  Before and after the measurements the candle weight was also
precisely measured and the paraffin consumption rate was estimated. Figure \ref{oxigen_konc} presents the results on oscillation frequency and the paraffin consumption rate as a function of oxygen concentration. 
The decreasing trend of the frequency and the growing paraffin consumption as a function of the increasing oxygen concentration was observed. It is interesting to note here that the oscillations frequencies of the flame measured inside the tube for the normal oxygen concentration of 21$\%$ are slightly lower than the values measured for 
 the 3 candle bundles outside the tube (Figure \ref{fig:frequency}a). This suggests that by limiting the space in which the combustion takes place will also influence the flickering frequency. We have also checked that the 
airflow inside the tube does not modify in a relevant manner the observed oscillation frequencies. Experimental results in case of two different incoming flow rates (200 l/h and 400l/h) are summarized in Table \ref{flow}.  From here we concluded that by doubling the flow rate there was no relevant effect on the observed oscillation frequency.    

\begin{table}[h!]
 \centering
\begin{tabular}{ |c|c|c|c|c|c|c| }
 \hline
 \multicolumn{1}{|c|}{flow rate}&\multicolumn{4}{|c|}{oscillation frequency [Hz]}&\multicolumn{1}{|c|}{$\langle f \rangle$ [Hz]}&\multicolumn{1}{|c|}{$\sigma(f)$ [Hz]}\\
 \hline
 200 l/h&9,89&10,31&10,03&10,54&10,19&0,29\\
 \hline
 400 l/h&10,5&9,9&10,58&10,49&10,36&0,31\\
  \hline
 \end{tabular}
 \caption{Influence of the incoming flow rate on the oscillation frequency inside the tube. Results for a bundle with three candles in triangular arrangement and 21\% oxygen concentration. Four different experiments were conducted at each flow-rate. The average frequency, $\langle f \rangle$, and the standard deviation, $\sigma (f)$, for the observed frequencies are given in the last two columns of the table.}
 \label{flow}
 \end{table}

\subsection*{\label{kiserletek_kollektiv} Experimental results on the collective behavior}

\begin{figure}[h]
    \centering
    \includegraphics[width=0.48\textwidth]{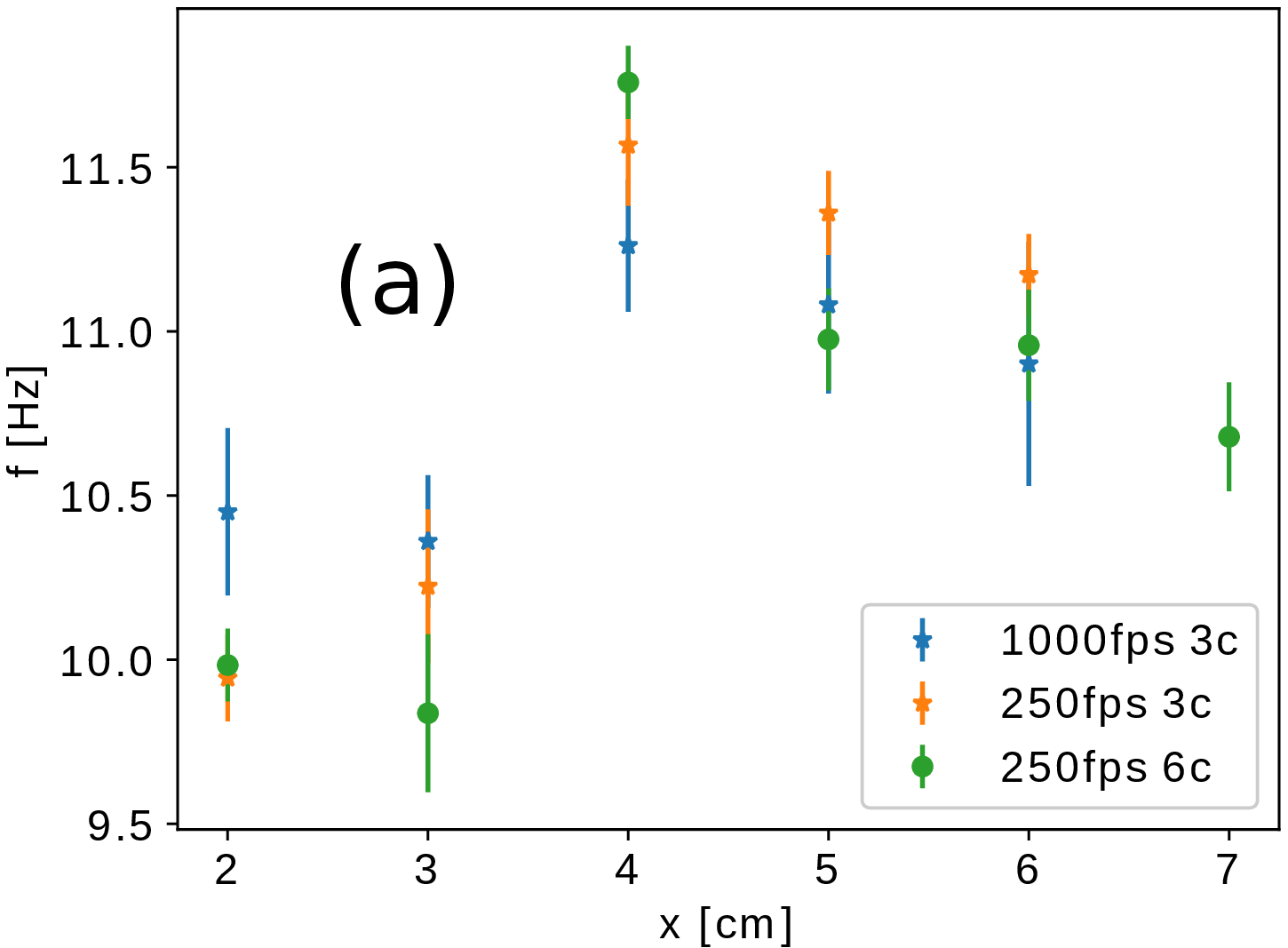}
    \includegraphics[width=0.48\textwidth]{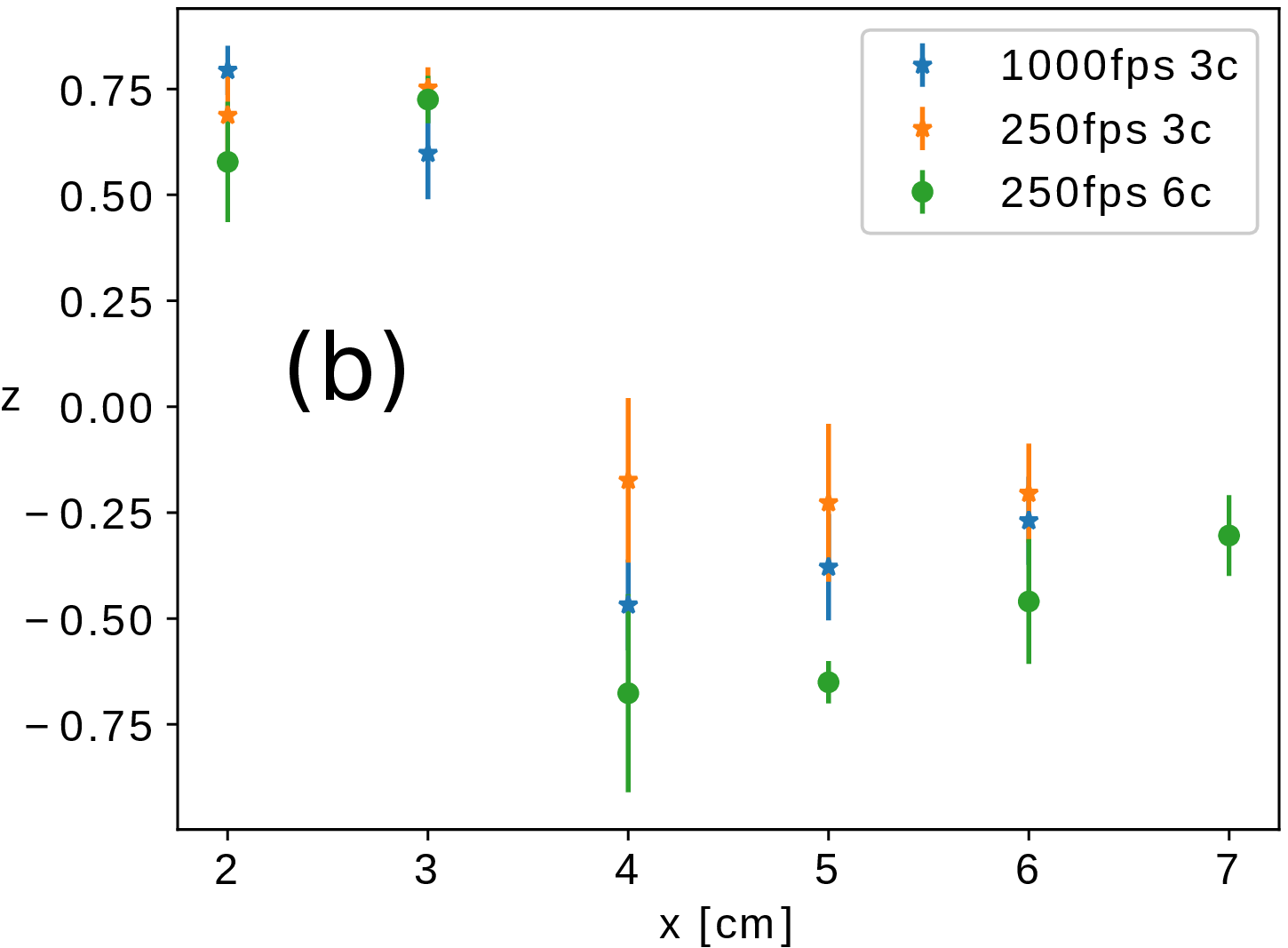}
    \caption{(a) Oscillation frequency of the triangular arrangement of triangular candle bundles containing 3 and 6 candles (denoted in the legend as 3c and 6c, respectively) as a function of the separation distance $x$. (b) Synchronization order-parameter calculated for the oscillations as a function of the separation distance. }
    \label{kollektiv_s}
\end{figure}

For  triangular arrangements the collective behavior of two bundles, with both 3 and 6 candles, was studied as function of the separation distance.
Movies recorded with the high-speed camera regarding the collective behavior of the flame of two candle bundles can be consulted on our YouTube playlist 
\cite{Attila3}. To quantify synchronization, we use the synchronization parameter $z$ borrowed from one of our earlier studies \cite{article}. In Supplementary Material, section B,  the calculation of $z$ is detailed. The value of this order 
parameter is 1 if the bundles are completely in-phase synchronized and it is -1 if they are completely anti-phase synchronized. The oscillation frequency  as a function of separation distance is plotted on  Fig \ref{kollektiv_s}a 
while the measured trend for the synchronization order parameter is presented in Fig \ref{kollektiv_s}b. 

It can be observed that for both bundles sizes the transition between the in-phase synchronization and counter-phase synchronization takes place between a separation distance from 3 to 4cm.  From the value of the measured $z$ order parameter we conclude that for bigger bundle sizes the counter-phase synchronization becomes more and more stable.

In order to verify the coupling mechanism through thermal radiation proposed by the Kitahata group \cite{Kitahata2009}, we 
replaced one candle bundle with a halogen bulb having the same radiation properties. The intensity of the current passing through the bulb was modulated so that its output radiation power and spectrum became similar to that of the flame. This radiation source was placed at 2cm distance from a triangular bundle containing 3 candles, and for different 
modulation frequencies of the bulb the collective behavior of the halogen bulb--candle bundle system was investigated.
 At such distance in-phase synchronization should be expected. Based on the recorded time series, the synchronization order-parameter was calculated and the results are presented in Table \ref{tab:sink_par}.  The values of $z$ 
 close to $0$ suggest no synchronization. In such a view one can thus
seriously question the validity of the coupling through the thermal radiation term in equations (\ref{eq:japan_coupled}).

\begin{table}[]
    \centering
    \begin{tabular}{|c|c|c|c|}
        \hline
       pulsation frequency&9.6 & 10.11 & 10.61\\
        \hline
        $z$&0.05& 0.06& 0.02\\
         \hline
        $\sigma(z)$&0.01&0.02 & 0.01\\
         \hline
    \end{tabular}
    \caption{Synchronization order-parameter, $z$, calculated for the collective behavior of a thermal radiation source oscillating at different frequencies and a triangular candle bundle containing 3 candles. We also indicate the
    value of the standard deviation $\sigma(z)$, estimated from 3 experiments in each case.}
    \label{tab:sink_par}
\end{table}

\section*{Improved dynamical model}

 In view of the presented experimental results it is obvious that the dynamical model of Kitahata et. al \cite{Kitahata2009} has to be reconsidered. 
The new model should successfully reproduces the detected trends in the frequency as a function of the candle numbers in the bundle and the oxygen concentration. 
On the other hand a successful model has to reproduce also the disappearance of the limit cycle for low and high candle numbers and oxygen concentrations as well. 
Also, the collective behavior has to be addressed by a realistic coupling and the detected synchronization order parameter and frequency has to be in agreement with the measured trends. 

Instead of the equation (\ref{eq:japan_eredeti}) we propose the following dynamical model:

\begin{equation}
 \begin{aligned}
C' N b(n_0) \frac{dT}{dt} &= \omega_1\left[-h'\,N\,b(n_0)\,(T-T_0)+\beta \, n\, N a(n_0) \, e^{-\frac{E}{RT}}\right]\\
\frac{dn}{dt} &= \omega_2\left[k\, (n_0-n)-N a\,(n_0) \,  n\, e^{-\frac{E}{RT}}\right]
\end{aligned}
\label{eq:modified}
\end{equation}

We neglected here the thermal radiation term, which was proven to be irrelevant for the collective behavior, and does
not influence  the limit cycle behavior either. The main difference relative to the original equations are the introduction of the $N \,b(n_0)$ and $N\,a(n_0)$ terms . 
The $N\,b(n_0)$ term describes the volume of the flame as a function of the candle numbers, $N$, and oxygen concentration, $n_0$. It is assumed that the volume of the flame increases linearly as a function of $N$. The dependence as a function of the used oxygen concentration, $n_0$, should have a more complicated functional form and this is why we use a general $b(n_0)$ kernel. Therefore if one denotes by $C'$ the volumetric heat capacity, the left side is the total thermal energy change inside the flame. We assume that the loss of heat by convection described by the first term on the right side is also governed by $N\,b(N_0)$ (this assumption is definitely more valid for the hollow and linear arrangements).  Instead of a constant $a$ term, governing the fuels supply rate we consider a
term depending again on both the candle number in the bundle and oxygen concentration: $N\,a(n_0)$. In the second equation for the dynamics of the $n$ variable the fuel supply rate has the same dependence. 
Let us also mention at this point that by taking into account Fick's law of diffusion, one can show that in reality $k$ can have also a weak dependence 
as a function of $N$, since $k\propto S/V$ where $S$ is the surface and $V$ the volume of the flame. Only for the hollow and linear arrangements one would get 
that $S\propto V$, and therefore no dependence as a function of $N$. We will accept in the following this assumption and consider $k$ as fixed. 

Before proceeding with these equations we just make a short note on the special case $T=T_0$ and $n=n_0$, which corresponds to a candle that is not burning. Naturally, this has to be a fix point of the system. Both the original equations (\ref{eq:japan_eredeti}) and our modified ones (\ref{eq:modified}) has the deficiency that this point will not be an exact fix point.  However, if we assume that the usual flame temperature is much bigger than $T_0$,  the  Boltzmann factor with the chemical activation barrier $E$
becomes very low for $T_0$. Therefore, the second terms on the right-hand side of equations (\ref{eq:modified}) will become negligible relative to the first ones and
as a result $T=T_0$ and $n=n_0$ becomes with a good approximation a fix point. In order to achieve the same situation in the original equations of Kitahata et. al
\cite{Kitahata2009} one has to add in the equation for the temperature variation the obviously missed +$\sigma T_0^4$ term.

Introducing again the non-dimensional  parameters from equations (\ref{eq:padim}) with the following modifications

\begin{equation}
     \varepsilon^{-1}=\frac{h'\omega_1}{C'\,k\,\omega_2};  \; a_u=\frac{\beta c \,n_0\, a(n_0) }{T_0 \,h' \,b(n_0)}e^{-c}; \, \;  a_v=\frac{N\,a(n_0)}{k}e^{-c}    
     \label{eq:ch1}
\end{equation}
we get the non-dimensional coupled dynamical equations:
\begin{equation}
\begin{aligned}
\frac{du}{d\tau} &= \frac{1}{\varepsilon}\left[-u+a_uve^{\left(\frac{uc}{c+u}\right)}\right]\\
\frac{dv}{d\tau} &= 1-v-a_vve^{\left(\frac{uc}{u+c}\right)}
\end{aligned}
\label{eq:sadim}
\end{equation} 

\begin{figure}
    \centering
    \includegraphics[width=0.45\textwidth]{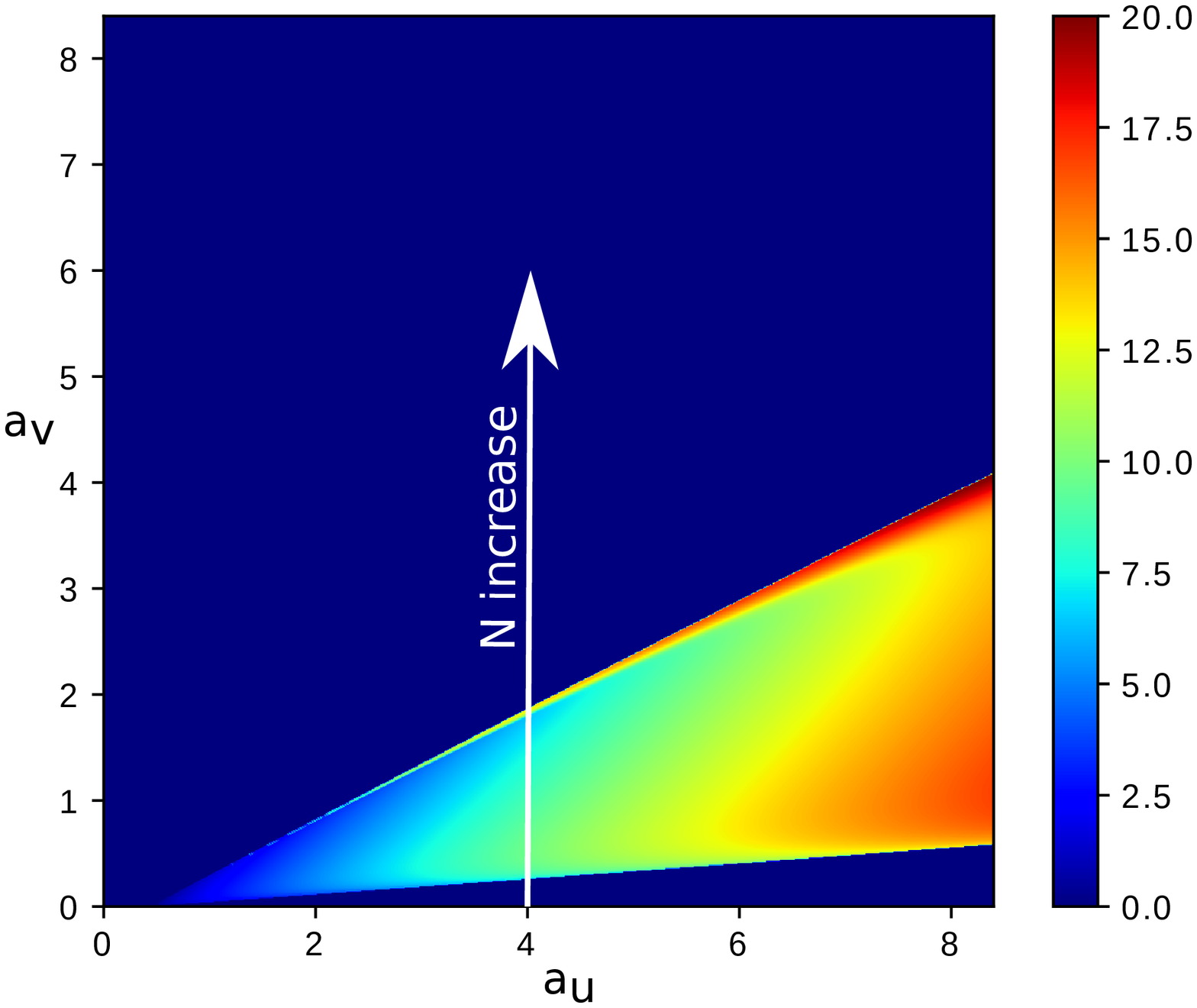}
    \includegraphics[width=0.45\textwidth]{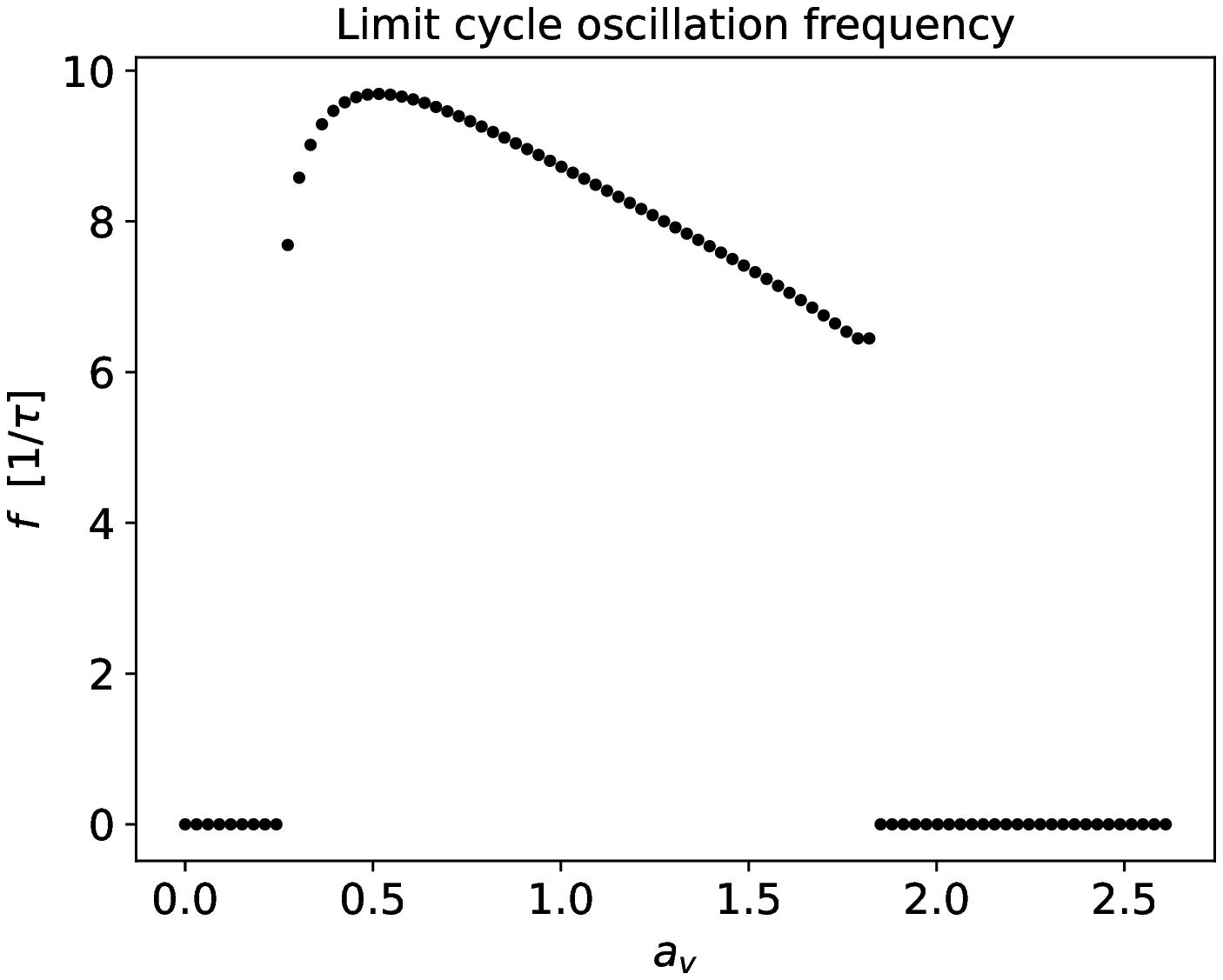}
    \caption{The figure from left shows the frequency of the flame oscillations in the $a_u-a_v$ parameter space illustrated with the color code from the legend. The vertical path suggested by the black line illustrates the path when one increases the number of candles in the bundle. In the dark violet region there is no limit-cycle and the system converges to a fix point. The other parameters were chosen as previously: $\varepsilon = 10^{-3}$, $c = 5.1$. The figure from right shows the oscillation frequency of the limit cycle on the white line direction in the figure from the left. }
    \label{fig:new_au-av}
\end{figure}

These equations are very similar to the original one, the main differences are however the relation between the $a_u$ and $a_v$ parameters influenced by the number of candles, $N$, and the oxygen concentration, $n_0$. Also, the last term due to thermal radiation is neglected. Similarly with the previous system (\ref{eq:japan_eredeti}), these new coupled dynamical equations allow for either a limit-cycle or a fix point. For the same parameters as in (\ref{eq:japan_eredeti}) in 
the left panel of Figure \ref{fig:new_au-av} we illustrate the frequency dependence in the $a_u-a_v$ parameter space. Again, in the violet region there is no limit-cycle, suggesting a stable flame. 
The form of the non-dimensional parameters $a_u$ and $a_v$ from equations (\ref{eq:ch1})  suggest that by increasing the 
number of candles in the bundle one follows a vertical path in the $a_u-a_v$ parameter space as it is indicated in
Figure  \ref{fig:new_au-av}. 

For example in the case of $a_u=4$ we get a region where there is a clear decreasing trend of the oscillation frequency as a function of $a_v$, as it is illustrated in the right panel of Figure \ref{fig:new_au-av}. For other $a_u$ values the behavior is rather 
similar, indicating a good agreement with our experimental observations.

The study of the system as a function of the used oxygen concentration $n_0$ is more complicated due to the 
unknown $a(n_0)$ and $b(n_0)$ kernel functions influencing both the $a_u$ and $a_v$ parameters. The $a(n_0)$
kernel in principal can be determined from the results presented in Figure \ref{oxigen_konc}. In experiments we observed that for $n_0< 0.17$ the oscillation stops, and for $n_0<0.15$ the flame also stops. The results presented in Figure \ref{fig:new_au-av} are in agreement with these observations. It suggests that there is a stable flame for low $a_v$ values, as observed in the experiments. Rising $n_0$ increases both the $a_u$ and $a_v$ values. Assuming the $b(n_0)$ increases in a more substantial manner than $a(n_0)$ we get that there is a possibility that for high $n_0$ values the system has again a fix point instead of the limit-cycle behavior. This was observed also in experiments for $n_0>0.9$ when the oscillation of the 
flames stopped.
 
For the collective behavior of two interacting flames we argued experimentally that the coupling 
through the thermal radiation is not realistic. Therefore we suggest a novel coupling based on the fact that the oscillation of each flame creates additional air flux perturbation, increasing the oxygen content around the adjacent flames. This hypothesis is embedded in the 
dynamical model by a coupling through the oxygen concentration variation as indicated in 
equations ($\ref{eq:ss}$) for two identical bundles $i,j\in\{1,2\}$.  The last term in the dynamical equations for the oxygen concentration in the flames results from a  simple argument based on ideal gas expansion and it is detailed in Supplementary Material, section C. According to this the oxygen concentration in one flame depends on the temperature change in the other. Any fluctuation in this temperature perturbs the air flow, increasing  the oxygen concentration in the nearby flame and the modulus indicates this symmetry. As detailed in Supplementary Material, section C, the effect will decrease as the inverse of the 
square of the distance. Denoting by  $\gamma$ a proportionality factor, the coupled equation system writes as: 
\begin{eqnarray}
C^{'}b(n_0) \frac{dT_i}{dt} =\omega_1 [h^{'}b(n_0)(T_0-T_i)+\beta a(n_0) n_i e^{ (-\frac{E}{RT_i})}] \nonumber \\
\frac{dn_i}{dt}=\omega_2[k(n_0-n_i)-N\, a(n_0) n_i e^{-\frac{E}{RT_i}}]+\left| \frac{dT_j}{dt}\right| \frac{\gamma}{x^2} \nonumber\\
\label{eq:ss}
\end{eqnarray}

\begin{figure}[h]
    \centering
    \includegraphics[width=0.45\textwidth]{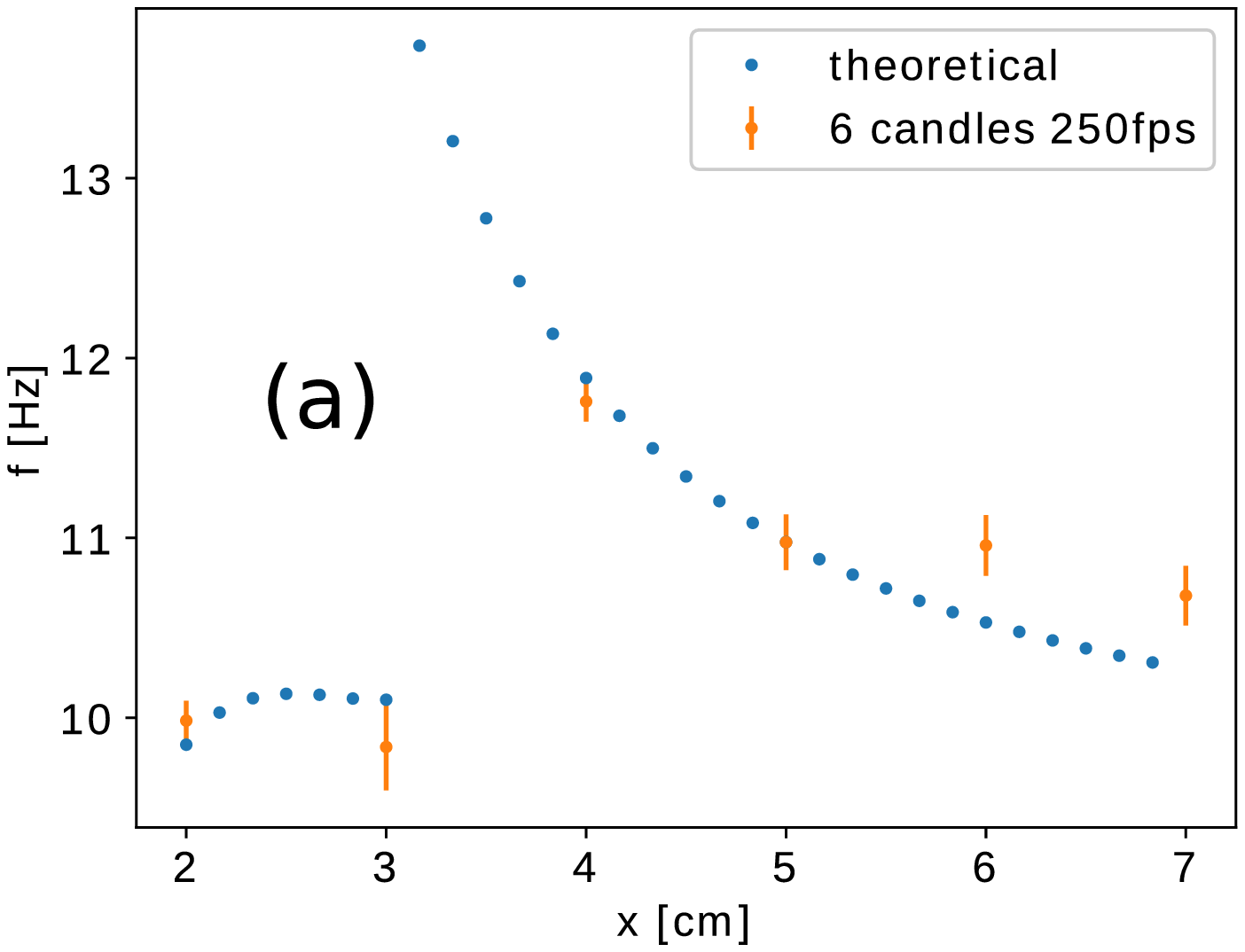}
    \includegraphics[width=0.45\textwidth]{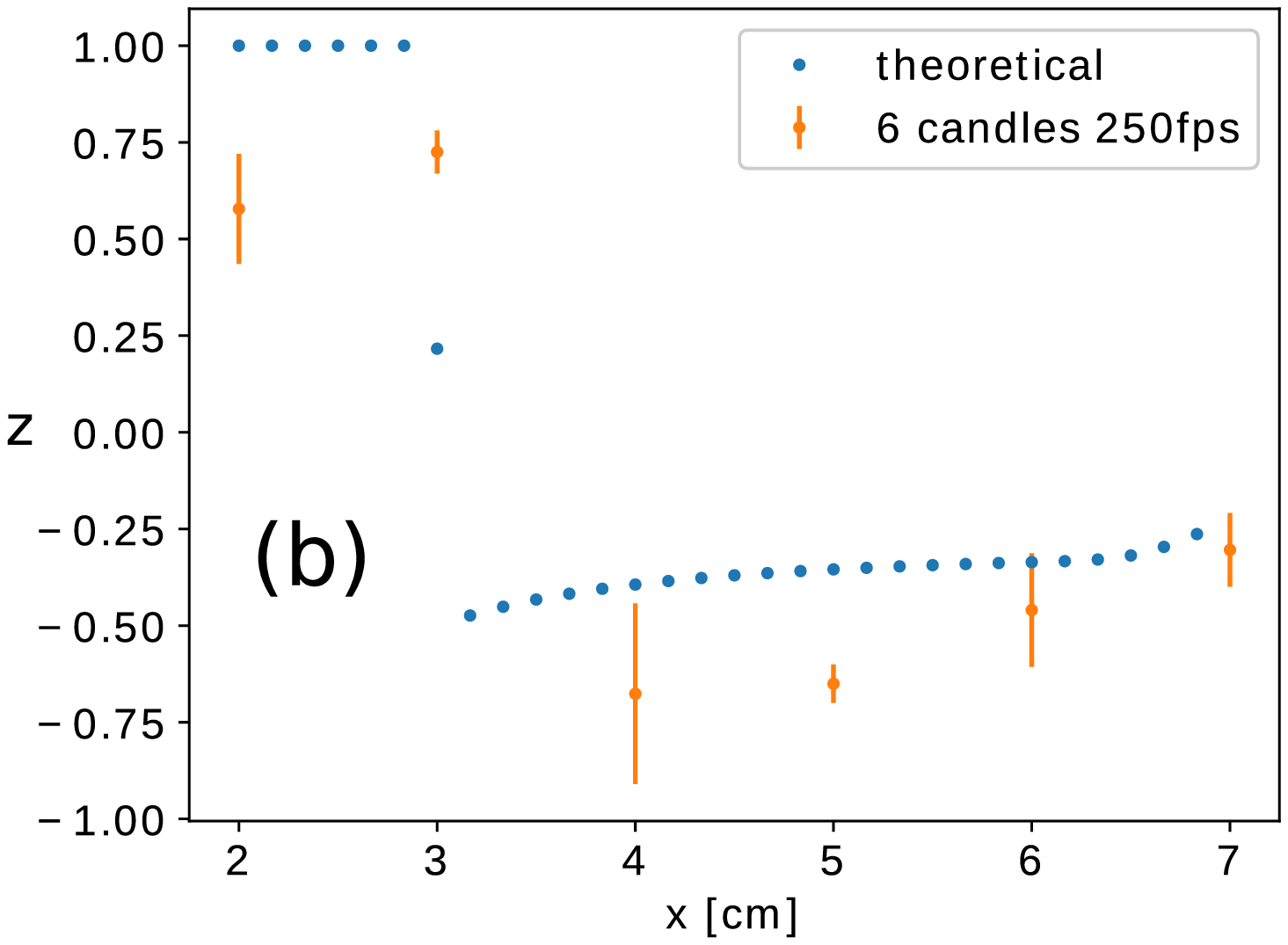}
    \caption{Experimentally and theoretically observed frequency (a), and synchronization order 
    parameter (b), for the collective behavior of two flames as a function of the separation distance $x$. For numerical calculations we used the parameters $\varepsilon = 10^{-3}$, $a_u = 10$, $a_v = 2$, $c = 5.1$, $\gamma=2.52\cdot 10^{-3}$ and  the time in numerics was adjusted to the experimental time-scale by $\tau=t \times 1.9$}
    \label{fig:exp_th}
\end{figure}

This new type of coupling reproduces excellently the observed in-phase and counter-phase synchronization and the frequency of the collective oscillations as function of the separation distance $x$. Results in this sense are compared in Fig. \ref{fig:exp_th}. For the previously used parameters ($\varepsilon=0.001$, $c=5.1$) and choosing $a_u=10$, $a_v=2$ and $\gamma=2.52\cdot 10^{-3}$ the theoretical results for the order parameter and the collective frequency are in excellent agreement if one rescales the time as $\tau=t\times 1.9$.  The $\gamma$ parameter governing the distance unit in numerical calculations was chosen properly, in order to ensure similarity between the collective behavior in the theoretical and experimental result. The transition region between the in-phase and anti-phase oscillations is clearly observable in the trends for the frequency and order-parameter. These results give again confidence in the new model and coupling mechanism.

\section*{Discussion and Conclusions}

The dynamical behavior of diffusion flames is a fascinating phenomena with many aspects still to explore. Here a specific example was investigated by considering the flame of candle bundles. In agreement with previous works on this system \cite{Kitahata2009,Forrester2015,Chen2019,Manoj2018,Ghosh2010,Okamoto2016} we observed experimentally the oscillation of the flames and their collective behavior. Our experimental investigation revealed however further interesting aspects which lead us to modify the dynamical equations used by Kitahata et. al \cite{Kitahata2009}. We found that the oscillation frequency of a single candle bundle decreases with the increasing candle number in the bundle for several arrangement type of the candles. For a given oxygen concentration in the air, there is both a lower and upper number of candle in the bundle for which the flame oscillates.  Similar results are valid when one changes the oxygen concentration of the air. Increasing the oxygen concentration will facilitate the appearance of the flickering, and for high enough oxygen concentration we observed also the flickering of one single candle. In order to explain the observed experimental facts we have shown that one can modify successfully the dynamical system proposed in \cite{Kitahata2009}. Similarly, we studied the synchronized oscillations of two 
candle bundle flames when they are placed at small distances. In agreement with previous observations we found in-phase synchronization for short distances and anti-phase synchronization for larger separation distances. As a function of the separation distance we measured the value of a synchronization order parameter and the common oscillation frequency.  The variation of both quantities suggests a discontinuous phase-transition at a given distance where the in-phase synchrony changes to the anti-phase synchrony. In order to accommodate these experimental results in the framework of the dynamical model used for explaining the oscillation of a single flame, we 
followed a similar method with Kitahata et. al. We first proved  that the coupling mechanism used in \cite{Kitahata2009} through a thermal radiation term is unrealistic, and suggested a mechanism via the air transport. It was found that our coupled system describes excellently all the observed collective phenomena, including the trends in the 
synchronization order parameter and common oscillation frequency. 

As always, when one reconsiders old and well-studied problems, new aspects of the phenomena are revealed. This was the case for the present study as well. Surprisingly, we learned that our experiments revealed new and interesting aspects of this fascinating phenomena. Connecting the oscillation and collective behavior of oscillation of diffusion flames with hydrodynamical instabilities and collective behavior of these would be also an interesting study in the future. Many pure hydrodynamical experiments 
reveal similar behavior \cite{Yuan1994}, suggesting that the presence of the flame is not necessary in order to obtain the discussed phenomena. Further experimental and theoretical studies might bring us closer to a more general understanding. Although such studies are  mainly interesting from the fundamental science perspective, one can also think of many practical applications where the discussed problems could be relevant: stabilization of flames and hydrodynamic flows,  inducing their desired  oscillations either by varying the parameters of the system or by synchronization, controlling many interacting flames, etc...

\section*{\bf Acknowledgment}

The work of Z.N. and B.S. was supported by the UEFISCDI research grants PN-III-P4-ID-PCCF-2016-0084 and PN-III-P1-1.1-PD-2019-0742.
A. G. acknowledges a STAR UBB excellence bursary from the Council for Research of the Babes-Bolyai University.

\section*{\bf Author contributions statement}

Conceptualization by Z.N.; Z.N. , A.G and Cs. P. conceived the experiments;  A.G. and R.T conducted the experiments;
Z.N. and B.S. conceived the models; A.G. analysed the results and models; Z. N. and A.G wrote the first version of the manuscript.  All authors reviewed the manuscript.

\section*{\bf Additional information}
\textbf{Competing interests:} the authors declare no competing interests.






 





\bibliography{candle_bib}









\end{document}